# A New Fault-Tolerant M-Network and its Analysis

R.N.Mohan  Pralhad Kulkarni and  Moon Ho Lee

*Abstract*-- This paper introduces a new class of efficient interconnection networks called as M-graphs, for large multiprocessor systems. The concept of the M-matrix and the M-graph is an extension of the $M_n$-Matrices and the $M_n$-graphs. We analyze these M-graphs to ascertain their suitability for large multiprocessor systems. An (p, N) M-graph consists of N nodes, where p is the degree of each node. The topology is found to possess many attractive features, prominent among them are its capability of maximal fault tolerance, higher density and constant diameter. It is found that these combinatorial structures exhibit some properties like symmetry and an interrelation with the nodes and degree of the concerned graphs, which can be effectively utilized for the purposes of inter connected networks. But many of the properties of these mathematical and graphical structures still remained unexplored and the present aim of this paper is to study and analyze some of the properties of these M-graphs and to explore their applications in networks and multi-processor systems.

*Index Terms*— Connectivity, fault-tolerance, interconnection networks, M-networks,  multiprocessor systems.

I. INTRODUCTION

Recent  developments in high performance multi-processor systems offer promising and powerful mechanisms for large-scale computation. The chances of failure of various components (nodes/links), however, also increase with the increase in the number of components. For this we adopt some recent structures namely the $M_n$-Matrices and the $M_n$-graphs, which were defined by Mohan in [1], and by extending this concepts, the M-matrix and the M-graph were defined by him in [2] recently, which we use for defining some new network systems. In fact the ability to

Manuscript received February 21, 2006.  The authors gratefully acknowledge the support received for this research work from Ministry of Information and Communication (MIC), under the IT Foreign Specialist Inviting Program (ITFSIP), supervised by the Institute of Information Technology Assessment (IITA), and the Chonbuk National University (CBNU). Our thanks are due to Raphael Favier, France, for his kind help in computer calculations.

R. N. Mohan is with the Division of Electronics and Information Engineering, Chonbuk National University, S. Korea, and on leave from the Sir. CRR Institute of Mathematics, Eluru- 534007, A.P., INDIA (telephone: +91-8656-232758, e-mail: mohan420914@yahoo.com.
P. T. Kulkarni is with the Division of Electronics and Information Engineering, Chonbuk National University, S. Korea, and on leave from the PES Institute of Technology, Bangalore-560085,  INDIA. e-mail: ptkul@ieee.org.
Moon Ho Lee is with the Division of Electronics and Information Engineering, Chonbuk National University, Jeonju-561756, S. Korea. Email: moomho@chonbuk.ac.kr.

tolerate failures greatly depends on the topological characteristics of the network [3]. Several topologies have been proposed and their fault-tolerance and fault-diagnosis capabilities have been analyzed [4]-[9]. All the proposed topologies have attractive features, as well as some inherent limitations. For example, a completely connected network might offer high fault-tolerant capability but costly to provide a direct communication path between each pair of processors. Alternatively, a loop structure is cost-effective but has poor fault-tolerant capability as well as large inter-node distances. Hypercube structure has nodal degree that increases with the network size, which necessitates additional expensive nodal interfaces (including transceivers). On the other hand, bounded degree networks shuffle-exchange and deBruijn networks have the advantage of requiring a fixed number of connections per node independent of the size of the network. Bounded degree networks do not allow the level of fault-tolerance to grow with the size of the network. It is of practical significance to design networks, which can achieve a desired level of fault-tolerance independent of the size of the network. Therefore it is important to design a network system having the feature of the simplicity of bounded degree networks as well as the versatility of unbounded degree networks such as the Hypercubes. The Hyper-deBruijn graphs are proposed [10] to accommodate this. Generally speaking, stochastic communication has a very good average case behavior with respect to latency and energy dissipation. In the worst possible case, the protocol will not deliver the packet to the destination; therefore this communication paradigm perhaps is better suited for applications, which can tolerate a small amount of information loss [11], [12].

These are some of the factors that motivated us for proposing a new class of networks. We have proposed a new family of networks called as M-networks, by considering the M-graphs.

And in this paper, we also analyze their properties for application in large multiprocessor systems. The present analysis reveals the interconnection complexity and the fault-tolerant capability of the proposed interconnection networks. The network reliability evaluation has become very important in the dimensioning and the design of telecommunication network. The emergence of wireless sensor networks has further increased the importance of fault-tolerance. In many parallel computer systems, processors are connected by interconnection networks. Such networks usually have regular topology and smaller degree to allow scalability. Popular instances of interconnection networks include arrays, meshes, trees, and the Hypercubes. A central issue in the design of interconnection networks is fault-tolerance as it is essential for a large parallel system to work properly even when some processors fail. The fault-tolerance of interconnection networks is more critical as the failure of one processor may affect the communication to other processors.

The rest of the paper has been organized as follows: Section II discusses the networks proposed earlier and their properties relevant to interconnection complexity and fault-tolerance. Section III gives the proposed network, the section IV discusses the analysis of this new network, and the conclusions have been drawn in Section V. Finally, in the Appendix, we give the 56-node computer aided M-network along with the list of successful communication paths for 12-node M-network, which are 56 in total number with varying hop-distances including distinct and alternate paths

## II. THE NETWORKS PROPOSED EARLIER

The N x M Manhattan Street Network (MSN) is a regular mesh structure of degree 2 with its opposite sides connected to form a torus. An unidirectional communication links connect its nodes into N rows and M columns, with adjacent row links and column links alternating in direction [13]. The (p,k) Shuffle Net can be constructed out of $N = kp^k$ nodes which are arranged in k columns of $p^k$ nodes each (where p,k = 1,2,3, ...), and the $k^{th}$ column is wrapped around to the first in a cylindrical fashion [14]. A $(\Delta, D)$-de Bruijn graph [15], $(\Delta \geq 2, \Delta \geq 2)$ is a directed graph with the set of nodes $(0, 1, 2, ..., \Delta-1\}^D$. The simplest form of the hypercube interconnection pattern is the binary hypercube [16]. A p-dimensional binary hypercube has $N = 2^p$ nodes, each of which have p neighbors. A node requires p transmitters and p receivers, and it employs one transmitter-receiver pair to communicate directly and bi-directionally with each of its p neighbors. Any node *i* with an arbitrary binary address will have as its neighbors those nodes whose binary address differ from node i's address in exactly one bit position. Its disadvantage is that the nodal degree increases logarithmically with *N*.

### III. THE PROPOSED NETWORK

This section presents some new interconnection networks and their properties, which are relevant to their application for multiprocessor systems. These networks can be represented as regular graphs (namely graphs in which all the nodes have the same constant degree d). The $M_n$-matrix, which was defined by Mohan in [1] (we quote hereunder all that needed comprehensively) as: By considering the equation $a_{ij} = d_i \otimes d_h d_j \mod n$ and suitably defining $d_i$, $d_h$, $d_j$ and $\otimes$ we get distinct $M_n$-matrices. That is w*hen n is a prime number, a symmetric $n \times n$ matrix obtained from* $[a_{ij}]$, *where* $a_{ij} = 1 + (i-1)(j-1) \mod n, \quad i, j = 1, 2, ..., n$. This has been used in the construction of certain graphs named as $M_n$-graph, which was defined as: *If given an* $M_n$-matrix and $C_k$ be its columns, which are numbered as 1,2,…,n and $a_{ij}$'s be the elements of it, then let $V_1 = \{C_k\}$, $V_2 = \{a_{ij}\}$ be the two sets of the vertices. Now there is an edge $α_{ijk}$ iff $a_{ij}$ is in $C_k$. Then the graph is ($V_1$, $V_2$, $α_{ijk}$), where i,j,k = 1,2,…,n is called as $M_n$-graph, which was used in an organizational set up as a net work in [1].

As an extension of these concepts, Mohan [1] and Mohan et al [2] defined some new (1,-1)-matrices called as M-matrices. The M-matrix of Type I had been defined as: *When n is a prime, consider the matrix of order n obtained by the equation $M_n = (a_{ij})$, where $a_{ij} = 1 + (i-1)(j-1)$ mod n, i, j = 1, 2,…,n. In the resulting matrix retain 1 as it is and substitute -1's for odd numbers and +1's for even numbers. (We can substitute 1 for odd numbers and -1 for even numbers; in that case change of sign occurs in its determinant). Let the resulting symmetric n×n matrix M be called as M-matrix of Type I.*

*Again by* considering the equation $a_{ij} = d_i \otimes d_h d_j$ mod n and by suitably defining $d_i$, $d_h$, $d_j$ and $\otimes$ we get *the M-matrix of the Type II, which is obtained by the equation $a_{ij} = (i.j)$ mod n+1, where n+1 is a prime number. In the resulting symmetric n×n matrix substitute 1 for even numbers and -1 for odd numbers and also for 1, (or 1 for odd numbers keeping the 1 in the matrix as 1 itself and -1 for even numbers). Then this resulting matrix M is called M-matrix of Type II.*

And these two types of the next generation matrices lead to corresponding graphs called as M-graphs. We find that the second type of graphs are more useful than the first as in the first type there are multiple connections between one pair of nodes (of different sets), and the maintenance of which becomes costly in network application point of view.

Thus as a further development of these concepts [2] a method of construction of M-graphs has been defined, which we use here for the construction of an M-network system. *In the given M-matrix of either type, replace -1's by 0's and let the resulting matrix be taken as the adjacency matrix of a graph, which is called as M-graph.* We give the concerned examples in the later sections. In our present discussions we use M-matrix of Type II only. Depending on that a (p,N) M-graph can be constructed out of N= 2n nodes where n+1, (this is for the sake of avoiding multiple connections between some pair of nodes of different sets) is a prime number > 3, and p = N/4 represents the degree of the graph. Because in the given M-matrix as n even in each row and in each column the number of +1's is n/2 and hence the degree of the concerned graphs will be n/2. Here N is the total number of nodes and n is the number of nodes in one set. As an example a (6, 24) M-network, which is drawn by a computer is shown below in figure 1.

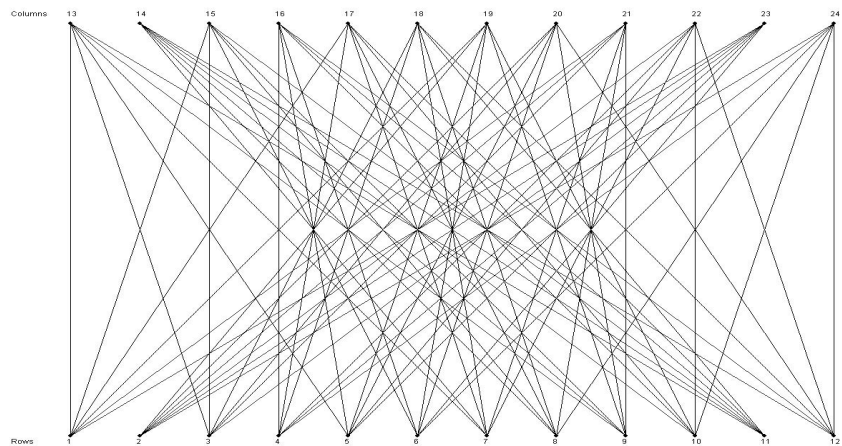

Fig. 1.  A  24-node M-network.

Let G = (V,E), where V is a set of vertices and E is a set of edges joining any two vertices, be called as graph. The number of edges passing through a vertex is called its valence/degree. In a graph, if the valence of each of its vertex is constant it is said to be regular. If the vertex set V has two complementary sets $V_1$, $V_2$ such that each edge of the graph has one end in $V_1$ and the other end in $V_2$, then the graph is called a bipartite graph. A graph is called k-*node-connected* if the removal of any subset of k-1 nodes leaves the graph connected, while there exists a subset of k nodes, whose removal disconnects the graph.

To explain the constructional method of this M-matrix of Type II, we give an example hereunder.

**Example.1.** Take n = 4, then n+1 = 5 a prime. Consider the equation $M = [a_{ij}] = (i.j)$ mod 5 for i, j = 1,2,3,4. We get the $M_n$ -matrix as follows:

$$\begin{bmatrix} 1 & 2 & 3 & 4 \\ 2 & 4 & 1 & 3 \\ 3 & 1 & 4 & 2 \\ 4 & 3 & 2 & 1 \end{bmatrix}.$$

Now by substituting 1 for even numbers and -1 for odd numbers and for 1 also in the above matrix we get the M-matrix of Type II, which is given below.

$$M = \begin{bmatrix} -1 & 1 & -1 & 1 \\ 1 & 1 & -1 & -1 \\ -1 & -1 & 1 & 1 \\ 1 & -1 & 1 & -1 \end{bmatrix}.$$

Now consider -1 are as 0's. We get (0,1)-matrix as

$$\begin{bmatrix} 0 & 1 & 0 & 1 \\ 1 & 1 & 0 & 0 \\ 0 & 0 & 1 & 1 \\ 1 & 0 & 1 & 0 \end{bmatrix}.$$

By considering this as the adjacency matrix of a graph, which is concerning to the given M-matrix, and we can draw the M-graph and can be treated it as an M-network.

**Proposition. 2.1.** The existence of an M-matrix of order n, when n+1 is a prime, implies the existence of a regular bipartite graph.

Explanation: From the (0,1) matrix that we have taken, considering that as an adjacency matrix with elements as one set and the columns as another set, if an element 'i' is in column 'j' then (i, j) will be an edge, otherwise not. Then we have $V_1 = n$ and $V_2 = n$, and its valence is $p = n/2$, since in the matrix the number of +1's and -1's each of them is n/2. Hence it is a regular bipartite graph.  

**Example 2.1.**

If we take n = 6 as n+1 = 7 a prime, we get an M-matrix as follows.

**Construction:** From the equation (i.j) mod 7, for i, j = 1, 2, 3,4,5,6, we get the $M_n$-matrix as

$$\begin{bmatrix} 1 & 2 & 3 & 4 & 5 & 6 \\ 2 & 4 & 6 & 1 & 3 & 5 \\ 3 & 6 & 2 & 5 & 1 & 4 \\ 4 & 1 & 5 & 2 & 6 & 3 \\ 5 & 3 & 1 & 6 & 4 & 2 \\ 6 & 5 & 4 & 3 & 2 & 1 \end{bmatrix}.$$

In this by changing the even numbers as -1's and odd numbers as +1's and keeping 1's in the matrix as it is, we get the M-matrix as given below.

$$\begin{bmatrix} 1 & -1 & 1 & -1 & 1 & -1 \\ -1 & -1 & -1 & 1 & 1 & 1 \\ 1 & -1 & -1 & 1 & 1 & -1 \\ -1 & 1 & 1 & -1 & -1 & 1 \\ 1 & 1 & 1 & -1 & -1 & -1 \\ -1 & 1 & -1 & 1 & -1 & 1 \end{bmatrix}.$$

In it by changing -1's as 0's we get

$$\begin{bmatrix} 1 & 0 & 1 & 0 & 1 & 0 \\ 0 & 0 & 0 & 1 & 1 & 1 \\ 1 & 0 & 0 & 1 & 1 & 0 \\ 0 & 1 & 1 & 0 & 0 & 1 \\ 1 & 1 & 1 & 0 & 0 & 0 \\ 0 & 1 & 0 & 1 & 0 & 1 \end{bmatrix}.$$

Now by taking this resulting binary matrix as the adjacent matrix of a graph, we get a regular bipartite graph, which is called as M-graph as given below.

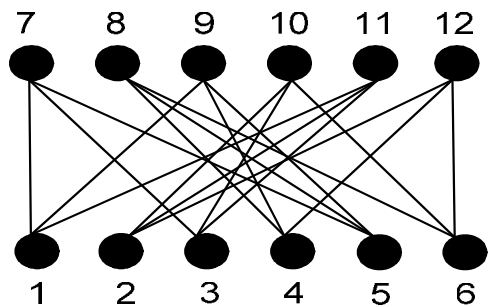

Fig. 2. A(3,12) M-graph.

We can generalize the method of generating these types of graphs by computer programming.

For the construction of M-networks with the required number of nodes and for finding n-hop paths in M-networks, Java program has been used. It has been found that there are 56 successful communication paths of different hop-distances, which consist of both distinct and alternate paths, establishing the fault tolerant capability of the above network. We propose to study some of the properties of the above M-graphs (M-networks) and find the application of it in the information and communication network systems. There are several important performance parameters for the system, such as message delays, routing algorithms, fault-tolerance, density, and mean hop distance, which are all depend on the topology, highly. These have been discussed in the next section.

## IV. ANALYSIS

Network reliability analysis is a difficult computational problem that has got attracted for further research at the interface of combinatorics, computer science and operations research. A central component of two-terminal reliability is the identification of minimum cutsets in a graph theoretic model of a network. The idea is to calculate the probability that at least one path exists between some selected pair of terminal nodes through the identification of all cutsets in the corresponding graph. Given a network with some small selected subset of vulnerable links, we would like to calculate the minimum number of network component failures that would have to occur so that all network communications involve at least one link in the vulnerable set.

Now we formulate the number of nodes in terms of hop distances in the given M-graph or (M-network) as given in the figure Fig.2. Here as it is a regular bipartite graph, we have the node set as two complementary sets, (1,2,3,4,5,6) and (7,8,9,10,11,12) and the degree is 3.

When in the regular bipartite graph, if the degree is p, from the source node $a_1$ say, then it has been connected to p nodes, namely $a_{p+1}$, $a_{p+2}$, $a_{p+3}$,…, $a_{2p}$. Here n+1 is a prime and p = n/2. Then in these p nodes, each of them have to contribute another p nodes and hence there will be $p^2$ nodes of the first set, which will occupy n-1 nodes, that is 2p-1 nodes. Hence $a_{p+1}$ goes to another p-1 nodes, $a_{p+2}$ goes to another p-1 nodes, …, $a_{2p}$ goes to another p-1 nodes and thus we have p(p-1) nodes that are to be occupied. But the columns concerned $a_{p+1}$, $a_{p+2}$, …, $a_{2p}$ will have some intersection numbers, when taken in pairs, for example columns concerned to $a_{p+1}$ and $a_{p+2}$ have some common elements. Thus the concerned columns have some common elements, accordingly that concerned elements will be repeating in these contributions and hence they are to be deleted. The same procedure is to be repeated until we arrive at only one element. In our present (3,12)-network system it turns out to be a 4-hop distance element. Hereunder we give an example to explain this procedure for that (3, 12) M-graph.

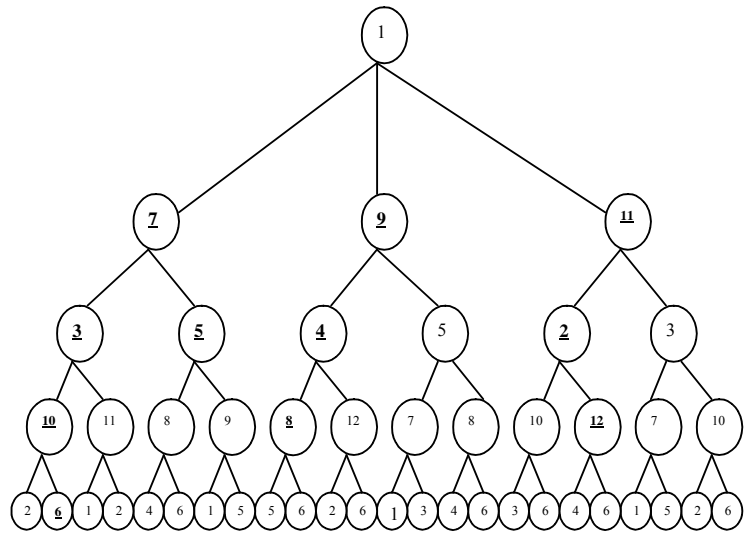

Fig. 3.   A Spanning Tree (With underlined nodes).

As an example, if node 1 has been taken as a source node and begin counting from it, as shown in Fig. 3, then connections will be from one set of nodes in a higher level to the nodes in a level immediately below it. In this figure all the possible node connections are indicated. However, the set of underlined-nodes, form a spanning tree. In this, the number of nodes that are one-hop away from the source node will be equal to the degree of the graph. For degree p = 3, the one-hop distance nodes are 7, 9, and 11. Then each of these three nodes will be connected to two nodes in the next level that are two hops away from the source. Hence there should be 9 such nodes. But there are only 4 new nodes namely 2, 3, 4, and 5, since the rest of the nodes have been covered already. Depending on the columns concerned to 7,9,11 in the adjacency matrix, their intersection numbers are: The columns 7 and 9 have node 5 as common, the columns 7 and 11 have node 3 as common, hence removing these common nodes we remain with 4 nodes namely 2,3,4,5, which are 2-hop distance nodes. These four nodes in turn should connect eight nodes at 3-hop distance in the next level. After deleting the repeated elements, we have nodes 8, 10, and 12 as the next nodes at 3-hop distance. This can be achieved by the concerned row/column intersection number in the adjacency matrix. In a similar way, deleting the elements that are repeated, we remain with 6, which is a 4-hop distance node in the fourth and final level. The node connectivity can further be explained with the following examples.

**Example 3.1**

n+1 = 5, p = 2, N = 8, and Diameter = 4.

**<u>With Node 1 as a Source</u>**

---------------------------------------------------------------------------------------------------

```
    Nodes  5   7                      ---- Level 1 (1 hop away there are 2 Nodes)

 ----------------------------------------------------------------------------------

    Nodes  3   2                      ---- Level 2 (2 hops away there are 2 Nodes)

 ----------------------------------------------------------------------------------

    Nodes  6   8                      ---- Level 3 (3 hops away there are 2 Nodes)

 ----------------------------------------------------------------------------------

    Nodes  4   4                      ---- Level 4 (4 hops away there is 1 Node)

 ----------------------------------------------------------------------------------
```

**d-hop distance nodes for (2,8) M-network**

**Example 3.2**

n+1 = 7, p = 3, N = 12, and Diameter = 4.

## With Node 1 as a Source

```
 ----------------------------------------------------------------------------------

    Nodes 7    9     11                ---- Level 1 (1 hop away there are 3 Nodes)

 ----------------------------------------------------------------------------------

    Nodes 3    5    4    5   2    3    ---- Level 2 (2 hops away there are 4 Nodes)

 ----------------------------------------------------------------------------------

    Nodes 10 11 8 9 8 12 7 8 10 12 7 10 ---- Level 3 (3 hops away there are 3 Nodes)

 ----------------------------------------------------------------------------------

    Nodes 4  5  6  2  4  6  2  3  6    ---- Level 4 (4 hops away there is 1 Node)
```

---------------------------------------------------------------------------------------------------------

**d-hop distance nodes for (3,12) M-network**

*the repeated elements at the same level and already occurred elements, are to be eliminated.

The results are summarized in Table 1. It is observed that any node in this network can be reached within a maximum of 4 hops independent of network size. Furthermore, the number of nodes that can be reached with smaller number of hops is higher than that with higher number of hops. This is the clear indication (4 hops) of the lower end-to-end delay.

| Sr No. | n+1 | n | N | p | Nodes 1-hop away | Nodes 2-hops away | Nodes 3-hops away | Nodes 4-hops away |
|---|---|---|---|---|---|---|---|---|
| 1 | 5 | 4 | 8 | 2 | 2 | 2 | 2 | 1 |
| 2 | 7 | 6 | 12 | 3 | 3 | 4 | 3 | 1 |
| 3 | 11 | 10 | 20 | 5 | 5 | 8 | 5 | 1 |
| 4 | 13 | 12 | 24 | 6 | 6 | 10 | 6 | 1 |
| 5 | 17 | 16 | 32 | 8 | 8 | 14 | 8 | 1 |
| 6 | 19 | 18 | 36 | 9 | 9 | 16 | 9 | 1 |

Table 1. The number of distinct hop-distance nodes in the networks of different sizes.

The reliability analysis of these networks in terms of successful communication paths and their fault-tolerance is given below. We have considered a (3, 12) M-network and this type of study can easily be extended to networks of any size. In this example it is considered that the node 1 as source node and the node 6 as destination node. We have computed various alternate paths between the source node and the destination node and based on this, the network fault-tolerance has been evaluated (see Appendix).

**Proposition.4.1.** The upper bound for the node-disjoint paths between any two distinct nodes x, y $\in \{N\}$ is given by $\left\langle \frac{N-2}{d-1} \right\rangle$, where $\langle \ \rangle$ indicates the greatest integer.

Proof: The maximum possible number of disjoint paths depends on the d, where there is a d-hop distance path. Now consider a d-hop distance path, which requires d+1 nodes including the source and destination nodes. By deleting these two nodes from N and d+1 then the possible number of distinct paths will be $\left\langle \frac{N-2}{d-1} \right\rangle$. □

Some of the main features of these M-networks are

**Mean hop distance**: The mean hop distance can be calculated by using the relation

$$\bar{h} = \sum_{i=1}^{D} i\, p_i, \qquad \ldots(1)$$

where D is the diameter, $p_i$ is the probability of i hops. For example, the mean hop distance for our 12- node M-network can be calculated using the data given in example 3.2, as below.

$$\bar{h} = \sum_{i=1}^{D} i\, p_i = 1\frac{3}{11} + 2\frac{4}{11} + 3\frac{3}{11} + 4\frac{1}{11} = \frac{24}{11} = 2.18 \qquad (2)$$

**Symmetry:** The proposed M-network is symmetrical. A node symmetric network has no distinguished node, that is, the view of the rest of the network it is the same from any node. Rings, fully connected networks, and hypercubes are all node symmetric, whereas trees and stars are not. When a network is node asymmetric, a distinguished node can become a communication- bottleneck.

**Density:** In these networks the majority of the nodes can be reached with the fewer hops from the source node. This is the clear indication of the higher density and the lower end-to-end delay.

**Connectivity:** The proposed networks have good connectivity and hence provide reliable communication through multiple paths between a given source-destination pair. The connectivity of these networks is found to be 0.5 (for fully connected case it is 1). Hence, we can say that these networks have good connectivity. The connectivity matrix for the present 12-node M-network is shown below.

The number of hops with which the nodes 1 to 12 are connected is shown in the matrix given below.

$$\begin{bmatrix} & 1 & 2 & 3 & 4 & 5 & 6 & 7 & 8 & 9 & 10 & 11 & 12 \\ 1 & 0 & 2 & 2 & 2 & 2 & 4 & 1 & 3 & 1 & 3 & 1 & 3 \\ 2 & 2 & 0 & 2 & 2 & 4 & 2 & 3 & 3 & 3 & 1 & 1 & 1 \\ 3 & 2 & 2 & 0 & 4 & 2 & 2 & 1 & 3 & 3 & 1 & 1 & 3 \\ 4 & 2 & 2 & 4 & 0 & 2 & 2 & 3 & 1 & 1 & 3 & 3 & 1 \\ 5 & 2 & 4 & 2 & 2 & 0 & 2 & 1 & 1 & 1 & 3 & 3 & 3 \\ 6 & 4 & 2 & 2 & 2 & 2 & 0 & 3 & 1 & 3 & 1 & 3 & 1 \\ 7 & 1 & 3 & 1 & 3 & 1 & 3 & 0 & 2 & 2 & 2 & 2 & 4 \\ 8 & 3 & 3 & 3 & 1 & 1 & 1 & 2 & 0 & 2 & 2 & 4 & 2 \\ 9 & 1 & 3 & 3 & 1 & 1 & 3 & 2 & 2 & 0 & 4 & 2 & 2 \\ 10 & 3 & 1 & 1 & 3 & 3 & 1 & 2 & 2 & 4 & 0 & 2 & 2 \\ 11 & 1 & 1 & 1 & 3 & 3 & 3 & 2 & 4 & 2 & 2 & 0 & 2 \\ 12 & 3 & 1 & 3 & 1 & 3 & 1 & 4 & 2 & 2 & 2 & 2 & 0 \end{bmatrix}.$$

A Comparison of various networks in terms of their characteristic parameters is shown in the Table 2. It is clear that the proposed M-network out-performs the other networks for application in interconnection network.

| Network | Degree | Size | Mean-Hop | Diameter | Scalability | Routing Complexity |
|---|---|---|---|---|---|---|
| Shufflenet [14] | p | $kp^k$ | $Log_p N$ | $2k-1$ | Poor | Very Low |
| deBruijn [15] | p | $p^n$ | $Log_p N$ | $Log_p N$ | Poor | Very Low |
| Binary Hypercube [16] | $log_2 N$ | $2^p$ | $Log_2 N$ | $Log_2 N$ | Poor | Very Low |
| Manhattan street network | 2 | Nr x Nc | $\sqrt{(Nr \times Nc)}$ | $\sqrt{(Nr \times Nc)}$ | Good | Low |
| M-network[13] | p | 4p | $2.18^*$ | 4 (constant) | Good | Very Low |

* For the 12-node M-network under consideration, see (1), (2)

Table 2. The comparison of various networks, given in terms of their characteristic parameters.

## V. CONCLUSIONS

In this paper, we have analyzed M-graphs for their fault-tolerant capability. These networks are shown to possess many attractive features, which make them competitive with other well-known architectures such as mesh, ShuffleNet, deBruijn, Hepercube and flip-trees etc. The maximum diameter of the M-network is found to be 4, independent of the network size. Hence, as the network size increases, these M-networks out-perform other known regular networks in terms of throughput and delay. They exhibit higher degree of fault-tolerance as these graphs have good connectivity and hence they provide a reliable communication system. However, their degree increases with the network size. These networks are found to be denser than many known multiprocessor architectures such as mesh, star, ring, the hypercube and its generalized topology [8]. A high density is desirable for interconnection networks, since such networks have relatively smaller communication delays. This study opens new tractable research directions. However, further work in these lines can be seen in a sequel to this paper, to appear shortly.

## ACKNOWLEDGMENT

The authors acknowledge the support received for this research work from Ministry of Information and Communication (MIC), under the IT Foreign Specialist Inviting Program (ITFSIP), supervised by the Institute of Information Technology Assessment (IITA), and the Chonbuk National University (CBNU). The authors are expressing their profoundest thanks to Prof. Moon Ho Lee, who facilitated them to work at Institute of Information and Communication of Chonbuk National University. Also thanks are due to Raphael Favier, France, for his kind help in computer calculations.

# APPENDIX I

The alternate paths between source-destination pair are computed to evaluate the network fault-tolerance. The results for 12-node M-network with node 1 as source and node 6 as destination are discussed below. We can compute all possible independent (node-disjoint) paths as well as alternate paths between given source-destination pair. For example, all possible independent (node-disjoint) paths are computed by considering node 1 as source node, in a 12-node M-network. From node 1 to node 2 there are 3/4 node-disjoint paths: (1,7,3,10,2 or 1,7,3,11,2), (1,9,4,12,2), (1,11,2). Similarly, between node 1 and node 7, there are three node disjoint paths: (1,7), (1,9,5,7), (1,11,3,7). This speaks of the good connectivity of the network. In any successful paths of s nodes and s-1 paths, it is trivial that if the remaining 2n-s nodes or nd-s+1 paths, where d is the degree of the graph, are potentially faulty, still the network system works successfully. In our 12-node network system, we have as follows:

There are 8, 4-hop successful paths and in each path 5 nodes and 4 edges are being used, which are given by (1,7,3,10,6), (1,7,5,8,6), (1,9,4,8,6), (1,9,4,12,6),(1,9,5,8,6), (1,11,2,10,6), (1,11,2,12,6), (1,11,3,10,6). Hence, even when the other 7 nodes and the other 14 edges fail, still the communication is possible. Therefore, in this case the network is said to be 7-node, 14-edge fault-tolerant.

There are 12, 6-hop successful paths, in each 7 nodes and 6 edges are used, which are given by (1,7,3,10,2,12,6), (1,7,3,11,2,10,6), (1,7,3,11,2,12,6),(1,7,5,8,4,12,6), (1,7,5,9,4,8,6), (1,7,5,9,4,12,6) ,(1,9,4,12,2,10,6), (1,9,5,7,3,10,6), (1,9,5,8,4,12,6), (1,11,2,12,4,8,6), (1,11,3,7,5,8,6), (1,11,3,10,2,12,6), Hence, even when the other 5 nodes and the other 12 edges fail, still the communication is possible. Therefore, in this case the network is said to be 5-node, 12-edge fault-tolerant.

There are 16, 8-hop successful paths, in each 9 nodes and 8 edges are used, which are given by (1,7,3,10,2,12,4,8,6), (1,7,3,11,2,12,4,8,6), (1,7,5,8,4,12,2,10,6), (1,7,5,9,4,12,2,10,6), (1,9,4,8,5,7,3,10,6), (1,9,4,12,2,11,3,10,6), (1,9,5,7,3,10,2,12,6), (1,9,5,7,3,11,2,10,6), (1,9,5,7,3,11,2,12,6), (1,9,5,8,4,12,2,10,6), (1,11,2,10,3,7,5,8,6),(1,11,2,12,4,9,5,8,6),(1,11,3,7,5,8,4,12,6),(1,11,3,7,5,9,4,12,6),(1,11,3,7,5,9,4,8,6), (1,11,3,10,2,12,4,8,6).

Hence, even when the other 3 nodes and the other 10 edges fail, still the communication is possible. Therefore, in this case the network is said to be 3-node, 10-edge fault-tolerant.

There are 20, 10-hop successful paths in each 11 nodes and 10 paths are used, which are given by

(1,7,3,10,2,12,4,9,5,8,6), (1,7,3,11,2,12,4,9,5,8,6),(1,7,5,8,4,12,2,11,3,10,6), (1,7,5,9,4,12,2,11,3,10,6)

(1,9,4,8,5,7,3,10,2,12,6), (1,9,4,8,5,7,3,11,2,10,6), (1,9,4,8,5,7,3,11,2,12,6), (1,9,4,12,2,10,3,7,5,8,6)

(1,9,4,12,2,11,3,7,5,8,6), (1,9,5,7,3,10,2,12,4,8,6), (1,9,5,7,3,11,2,12,4,8,6), (1,9,5,8,4,12,2,11,3,10,6),

(1,11,2,10,3,7,5,8,4,12,6), (1,11,2,10,3,7,5,9,4,8,6), (1,11,2,10,3,7,5,9,4,12,6), (1,11,2,12,4,8,5,7,3,10,6),

(1,11,2,12,4,9,5,7,3,10,6), (1,11,3,7,5,8,4,12,2,10,6), (1,11,3,7,5,9,4,12,2,10,6), (1,11,3,10,2,12,4,9,5,8,6).

Hence, even when the other 1 node and the other 8 edges fail, still the communication is possible. Therefore, in this case the network is said to be 1-node and 8-edge fault-tolerant network.

There are 56 total successful communication paths, having both alternate and distinct paths, besides having faulty nodes and edges, showing the possibility of successful communication system. This reveals that how best fault-tolerant the network is.

As an example for a higher node M-network, the computer generated 56-node M-graph is shown below.

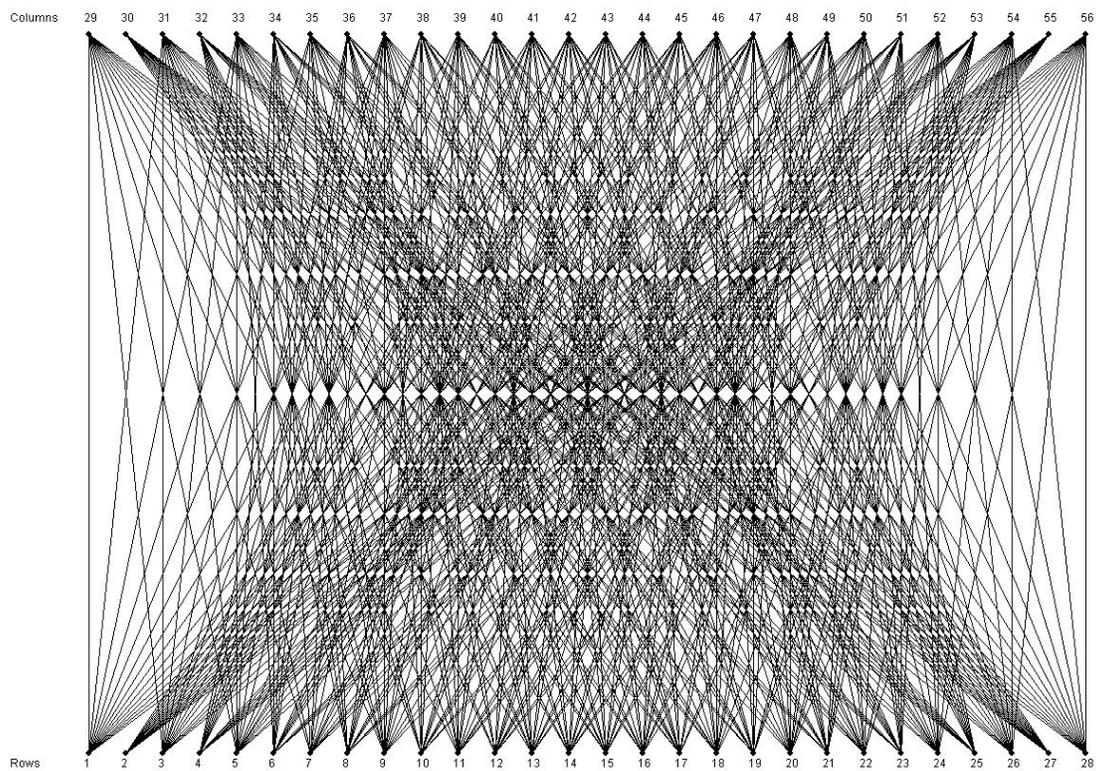

**56-node M-network system.**